\documentclass[twocolumn,showpacs,prb,amsfonts,amsmath,
               amssymb,floatfix]{revtex4}

\usepackage{graphicx}
\usepackage{dcolumn}
\usepackage{bm}

\begin{document}   
\preprint{cond-mat/0110430}
\title{Electronic structure of the {\it M}O oxides ({\it M}=Mg, Ca, Ti,
V) in the GW approximation}
\author{Atsushi Yamasaki}
\email{yama@coral.t.u-tokyo.ac.jp}
\author{Takeo Fujiwara}
\email{fujiwara@coral.t.u-tokyo.ac.jp}
\homepage{http://fujimac.t.u-tokyo.ac.jp/FujiwaraLab/index.html}
\affiliation{Department of Applied Physics, University of Tokyo,   
Bunkyo-ku, Tokyo 113-8656, Japan}   
\date{published 13 December 2002}
\begin{abstract}
The quasiparticle band structures of nonmagnetic monoxides, 
{\it M}O ({\it M}=Mg, Ca, Ti, and  V), are calculated by the GW
approximation. 
The band gap and the width of occupied oxygen 2$p$ states 
in insulating MgO and CaO agree with experimental observation.
In metallic TiO and VO, conduction bands originated from 
metal 3$d$ states become narrower. 
Then the partial densities of transition metal $e_g$ 
and $t_{2g}$ states show an enhanced dip between the two. 
The effects of static screening and dynamical correlation are discussed
in detail in comparison with the results of the Hartree-Fock
approximation and the static Coulomb hole plus screened exchange
approximation. 
The $d$-$d$ Coulomb interaction is shown to be very much reduced
by on-site and off-site $d$-electron screening in TiO and VO. 
The dielectric function and the energy loss spectrum are also 
presented and discussed in detail. 
\end{abstract} 
\pacs{71.10.-w, 71.15.-m, 71.20.Be}
\keywords{GW}
\maketitle 

\section{Introduction} \label{sec1:intro}
The first-principle electronic structure calculation, 
based on the local-density approximation (LDA) within the 
density-functional theory (DFT),~\cite{LDA}
has had great success in theoretical investigations of the ground state
property of condensed matters. 
However, in many LDA calculations, 
the band gap is unsatisfactory underestimated in insulators 
and semiconductors.
In the electronic structure of 3$d$ transition metals 
calculated by the LDA,
the occupied 3$d$ band width is too broad, the exchange splitting is
overestimated and the satellite structure in the x-ray photoemission
spectroscopy (XPS) does not appear.
These quantities are associated with quasiparticle properties.
The transition metal monoxides {\it T}O ({\it T}=Mn, Fe, Co, and Ni) are
wide-gap antiferromagnetic insulators in experiments, 
but FeO and CoO appear as metals, and MnO and NiO as small-gap
semiconductors in the LDA.~\cite{LDA-TMO}
The magnetic moments of {\it T}O are  underestimated by the LDA.

The electronic structure in strongly correlated electron systems
has been highly challenging  field.
The self-Coulomb and self-exchange interactions almost 
cancel with each other in many cases of the LDA but still not 
completely in {\it T}O.~\cite{GGA} 
The self-interaction corrected LDA (SIC-LDA) has been
applied to transition metal monoxides.~\cite{SIC2,SIC3}
It gives satisfactory results for the band gap and magnetic
moments in the transition metal monoxides {\it T}O (Tr=Mn, Fe, Co, and
Ni) except VO.
The LDA+U method~\cite{LDA+U1,LDA+U2,LDA+U3} works reasonably well for
the Mott-Hubbard insulators or rare earth metal compounds where the 
3$d$ or 4$f$ bands are partially filled. 
The LDA+U method is a static limit of the GW approximation
(GWA).~\cite{LDA+U3}
However, application of the LDA+U method to early
transition metal monoxides TiO and VO would lead to
antiferromagnetic insulators rather 
than paramagnetic metals.~\cite{LDA+U1}

The GWA is the first term approximation of the many-body perturbation
series for the Green function, and 
is a simple but excellent method  to calculate 
the self-energy with the dynamical correlation.~\cite{GW1,GW2} 
On the other hand, the HF approximation
overestimates the band gap in insulators and semiconductors 
because of the lack of the screening terms and
gives zero density of states at the Fermi level in the electron gas.
In the GWA, there is no self-interaction because the exchange term 
of the GWA is of the Hartree-Fock type.
The dynamical correlation effect in the GWA is treated by the
random-phase approximation (RPA).
In the GWA, the dynamical effects give rise to important features that
are not accounted for by the HF approximation, for example, damping of 
quasiparticles and characteristic structure in the spectral function.
The screening effects recover the finite density of states at the Fermi
level.
The band gap is corrected and reduced in the GWA by the dynamical
screening effects
in comparison with the one of the HF approximation.
The dynamical correlation effect is important
for the band width in the transition metals.~\cite{QP} 
We could not expect improvement in the GWA for the exchange splitting
and satellite structure because it needs higher order diagrams
(e.g., vertex corrections) for electron-electron and hole-hole scattering
processes.~\cite{Niexp}

The GWA has been applied to several real systems, 
for example, simple metals,~\cite{GW-SimpleMetal}
semiconductors and insulators,~\cite{GW-semicon1,GW-semicon2,GW-MgO}
transition metals,~\cite{GW-TM} and 
transition metal monoxides.~\cite{GW-TMO}
In simple metals and semiconductors,
the GWA can be formulated with plane wave basis set 
based on the pseudopotential method and  
the plasmon pole approximation has been often 
used for calculating dielectric function.~\cite{GW-semicon1}
In contrast, transition metals and their compounds have strong atomic
potentials and the GWA 
based on the pseudopotential method cannot be applied 
because of restriction of the number of plane waves.
It is essentially important to include core electrons in many cases
and, moreover, 
the plasmon pole approximation is not applicable to the dielectric
function for transition metals and their compounds 
because there is no isolated plasmon pole. 
Therefore, all-electron calculation and no plasmon-pole approximation in 
the dielectric function are required for transition metals and
their compounds. 

The GW calculation of Ni,
based on the linearized augmented-plane-wave (LAPW) basis and 
no approximation in the dielectric function, 
gives results in good agreement with experiments.~\cite{GW-TM} 
The modeling of the screened interaction in the GWA
is able to explain satisfactorily photoemission spectra in
MnO,~\cite{mGW-TMO1} NiO,~\cite{mGW-TMO2} and VO$_2$.~\cite{mGW-TMO3}
In the pioneering work of the GW calculation 
of NiO by Aryasetiawan,
the band gap and magnetic moment agree quite well with experimental
values.~\cite{GW-TMO}
In the analysis of photoemission spectroscopy by the cluster
configuration-interaction theory, 
NiO is assigned as a charge-transfer type insulator,~\cite{NiO-CI}
but in the LDA it is classified as a
Mott-Hubbard type insulator because the transition metal 3$d$ bands
appear at higher energy than oxygen 2$p$ bands.~\cite{LDA-TMO}
The relative position of the oxygen 2$p$ states of the GWA is also
improved in comparison with that of the LDA.  
However, the top of the valence bands in the GWA is mainly Ni 3$d$
states and NiO could not be a charge-transfer type insulator. 

In most GW calculations, full self-consistency is not carried out.
The self-consistency in the GW calculations can maintain the
conservation laws of particle number, energy, and
momentum.~\cite{Conserve} 
However such treatment cannot guarantee better agreement with
experimental results.~\cite{GW-SCF}

Because it is derived from many-body perturbation theory,
the GWA is applicable to wide variety of classes of systems,
and should be tested in various systems.
In this paper, the GW method based on the linear muffin-tin orbital 
(LMTO) method~\cite{OKA} is applied to several non-magnetic oxides, 
insulating MgO, CaO and metallic TiO, VO.
Only the electronic structure in MgO was calculated by the GWA 
before.~\cite{GW-MgO}
The insulators MgO and CaO are typical simple oxides, 
whose insulating band gap is underestimated by the LDA.
TiO and VO are nonmagnetic metallic oxides.
In the LDA their characteristics are very similar. 
We will discuss the electronic band structure and its systematic 
variation for MgO, CaO, TiO, and VO, concerning 
with the $E$-${\bf k}$ relation, 
band gap, band width, the density of states, plasmon frequencies, 
structures in the energy loss spectrum,  and other properties. 
We will also discuss the correlation effects, including the 
strength of the Coulomb interaction with $d$-$d$ screening effects.

The paper is organized as follows. 
A simple review on the theoretical framework 
will be given in Sec.~\ref{sec2:thory}. 
The results for these systems are presented  and discussed in
Sec.~\ref{sec3:result}. 
The dielectric function and the energy loss spectrum 
are also discussed.
Finally, in Sec.~\ref{sec4:summary} we present our summary.

\begin{table*}
\caption{The band gap $E_G$, the width of valence band $W_{v}$, 
         and the position of the O $s$ band $E_{O_s}$ 
         in insulating MgO and CaO.
         The unit is (eV).}
\begin{ruledtabular}
\begin{tabular}[t]{lccccccccccc}
          & \multicolumn{5}{c}{MgO}&         
          & \multicolumn{5}{c}{CaO}\\ 
          & LDA     & HF   & COHSEX    & GWA        & expt. &
          & LDA     & HF   & COHSEX    & GWA        & expt.   \\ \hline
$E_G$     & 5.2     & 18.4,\footnotemark[1]  16.5,\footnotemark[2]
            17.6\footnotemark[3]
          & 9.6     & 8.2    & 7.8\footnotemark[4]&
          & 3.65    & 15.9\footnotemark[3]
          & 7.7     & 6.64   & 7.1\footnotemark[4]\\
$W_{v}$   & 5.0     & 10.4,\footnotemark[1]  5.8,\footnotemark[2]
             7.64\footnotemark[3]
          & 5.6     & 5.0    & 5.0--6.0\footnotemark[4]&
          & 2.8     & 3.43\footnotemark[3]
          & 3.4     & 2.9    &        \\
$E_{O_s}$ & $\sim$ 16 & $\sim$ 40,\footnotemark[1] $\sim$ 24\footnotemark[3]
          & $\sim$ 19 & $\sim$ 17  & 18--21\footnotemark[5]&
          & $\sim$ 15 &   & $\sim$ 17 & $\sim$ 16  &       \\
\end{tabular}
\end{ruledtabular}
\footnotetext[1]{Reference~\onlinecite{MgO-HF1}.}
\footnotetext[2]{Reference~\onlinecite{MgO-HF2}.}
\footnotetext[3]{(Uncorrelated) HF results in Ref.~\onlinecite{HFcorr}. 
                 Correlated HF results are $E_G$=8.21~eV in MgO and
                                           $E_G$=7.74~eV in CaO.}
\footnotetext[4]{Reference~\onlinecite{MgO-CaO_exp}.}
\footnotetext[5]{Reference~\onlinecite{MgO_exp2}.}
\label{MgOCaOtbl}
\end{table*}%

\section{Theory} \label{sec2:thory}
\subsection{GW approximation}
When effects of dynamical correlation
are included, the self-energy is expressed formally as a series
expansion of dynamically screened interaction. 
The GWA replaces the self-energy by the lowest order term of 
the expansion as $\Sigma(1,2) = i G(1,2) W(1,2)$.
$G$ is the one particle Green function and
the dynamically screened interaction $W$ is defined by
\begin{align}
 W(1,2) &= \int d(3) \varepsilon^{-1}(1,3) v(3,2) , \label{dyn-int}\\
 &= v(1,2) + \int d(34) v(1,3) \chi^0(3,4) W(4,2) , \label{dyn-int2}\\
 &= v(1,2) + W^c(1,2) , \label{dyn-int3}
\end{align}
where $\varepsilon$ is the  dynamical dielectric function, $v$
is the bare Coulomb potential and $\chi^0$ is the irreducible
polarization function $\chi^0(1,2) = -i G(1,2)G(2,1)$.
Here we use an abbreviated notation 
$(1)=({\bf r}_1,\sigma_1, t_1)$ and
$v(1,2)=v({\bf r}_1,{\bf r}_2) \delta(t_1-t_2)$.
Equation (\ref{dyn-int2}) can be treated by the RPA.

We adopt the LDA Hamiltonian to be the unperturbed one
$H^0$,  
\begin{equation}
 H^0=T+V^H+V^{\rm xc}_{\rm LDA} \label{LDA-hamiltonian} .
\end{equation}
Here $T$ is the kinetic energy, $V^H$ is the Hartree potential,
and $V^{\rm xc}_{\rm LDA}$ is the exchange-correlation potential in the
LDA. 
We presume the wave functions $\{ \psi_{{\bf k}n}({\bf r})\}$ 
of the LDA to be a reasonably good starting wave functions, 
though this assumption should be carefully investigated in detail. 
Then the self-energy can be written by three terms as
\begin{equation}
 \Delta\Sigma = \Sigma^x + \Sigma^c - V^{\rm xc}_{\rm LDA} ,
\end{equation}
where $\Sigma^x (= i G v)$ is the exchange part (the Fock term) and 
$\Sigma^c (= i G W^c)$ is the dynamical correlation part. 
The quasiparticle energy is given as
\begin{equation}
 E_{{\bf k}n} = \epsilon_{{\bf k}n} + Z_{{\bf k}n} 
  \Delta\Sigma_{{\bf k}n} (\epsilon_{{\bf k}n}) ,
\end{equation}
where $\epsilon_{{\bf k}n}$ is the LDA eigenvalue. 
The self-energy $\Delta\Sigma_{{\bf k}n}$ and 
the renormalization factor $Z_{{\bf k}n}$, the weight of the
spectral function, are given by 
\begin{align}
 \Delta\Sigma_{{\bf k}n} (\epsilon_{{\bf k}n}) &= 
  \langle \psi_{{\bf k}n} | \Sigma^x + \Sigma^c(\epsilon_{{\bf k}n}) -
  V^{\rm xc}_{\rm LDA} | \psi_{{\bf k}n} \rangle , \\
 Z_{{\bf k}n} &= 
  \left[ 1 - 
   \left. \frac{\partial \Delta\Sigma_{{\bf k}n}(\omega)}
    {\partial\omega} \right|_{\omega=\epsilon_{{\bf k}n}} \right]^{-1} .
\end{align}
The renormalization factor $Z_{{\bf k}n}$ is a measure of 
the occupation number and should equal to the discontinuity of
occupation number at the Fermi energy. 
Therefore it should satisfy the condition $Z_{{\bf k}n} \leq 1$.
In the present work we perform one iteration calculation without
self-consistency.

In order to know the difference of the dynamical and static
screening effects, we also apply
the static Coulomb hole plus screened exchange (COHSEX) approximation
for the self-energy by Hedin.~\cite{GW1}
The static COHSEX approximation can be obtained by neglecting energy
dependence of the self-energy. 
Then the self-energy is expressed as
\begin{align}
 &\Sigma_{\rm SEX}({\bf r},{\bf r}') = - \sum_{{\bf k}n}^{\rm occ}
 \psi_{{\bf k}n}({\bf r})\psi_{{\bf k}n}^{*}({\bf r}')
 W({\bf r},{\bf r}';\omega\!=\!0) , \\
 &\Sigma_{\rm COH}({\bf r},{\bf r}') = \frac{1}{2} 
 \delta({\bf r}\!-\!{\bf r}')
 \left[ W({\bf r},{\bf r}';\omega\!=\!0) - v({\bf r},{\bf r}') \right]
 ,
\end{align}
where $\Sigma_{\rm SEX}$ is the screened-exchange term and 
$\Sigma_{\rm COH}$ is the Coulomb-hole term which becomes the local
potential in the static approximation.
The self-energy matrix element of $W({\bf r},{\bf r}^\prime: \omega=0)$
in the $d$ orbitals is the Coulomb interaction including 
the $d$-$d$ screening, which will be discussed in the following section. 

\subsection{LMTO minimal basis set and 
choice of LMTO parameters}\label{sec2-b:LMTO}
Because the plane wave basis becomes very costly 
for systems containing 3$d$ or 4$f$ electrons, 
the LMTO method~\cite{OKA} is more appropriate.
We use the LMTO basis set $\chi_{{\bf R}L\nu}({\bf r})$ within the
atomic sphere approximation (ASA) for the LDA calculation.
Here $L$ is angular momentum $L$ = $(l,~m)$.
The LMTO can be expanded by the muffin-tin orbital 
$\phi_{{\bf R}L\nu} ({\bf r})$ and it's energy derivative 
$\dot \phi_{{\bf R}L\nu} ({\bf r})$.
The product basis formalism~\cite{pd-basis} is used and proved to be
very efficient.
Computational details and numerical technique are shown in 
Appendixes~\ref{App.A} and ~\ref{App.B}. 

\begin{figure*}
\begin{center}
 \includegraphics[width=75mm,clip]{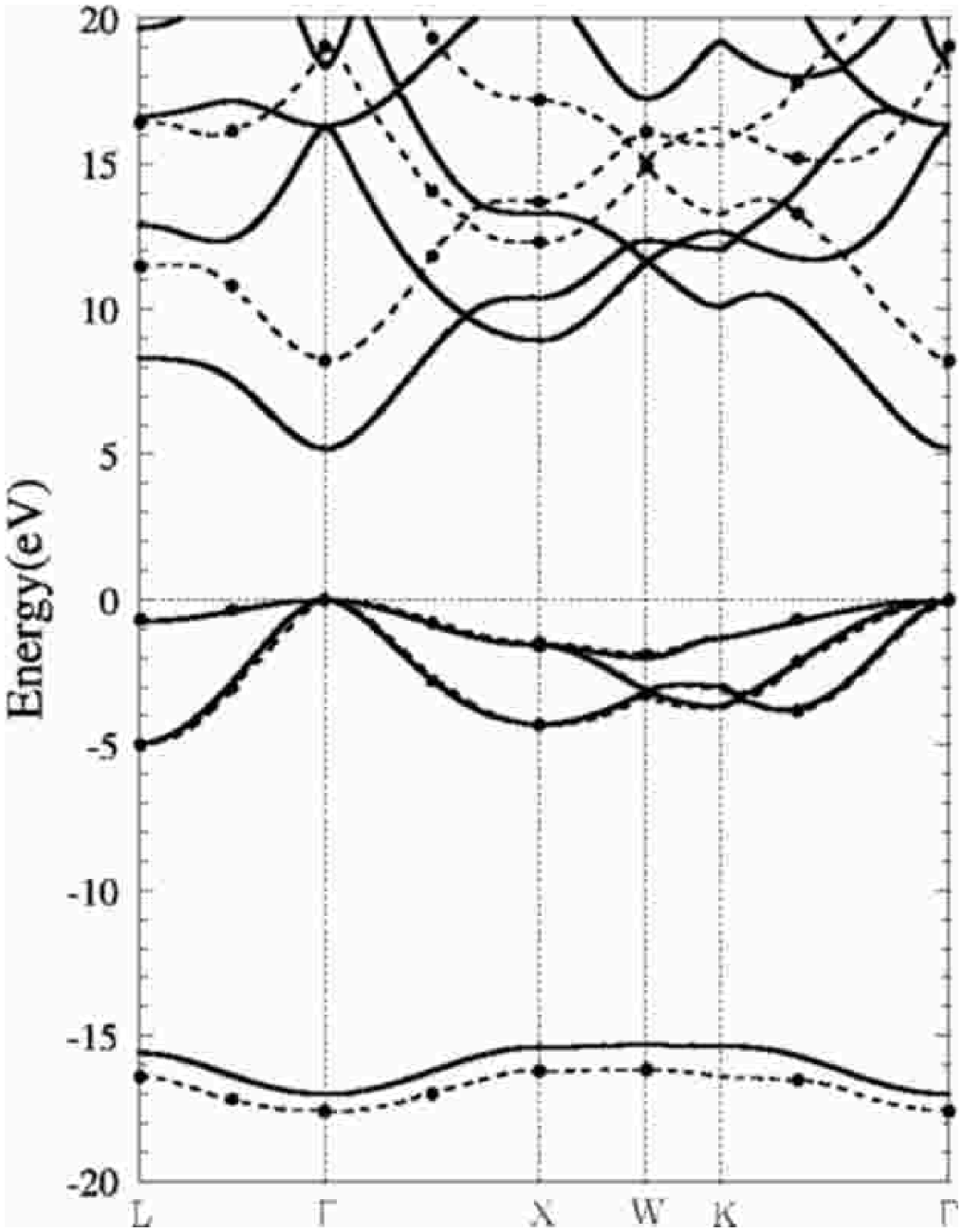}
 \hspace{5mm}
 \includegraphics[width=75mm,clip]{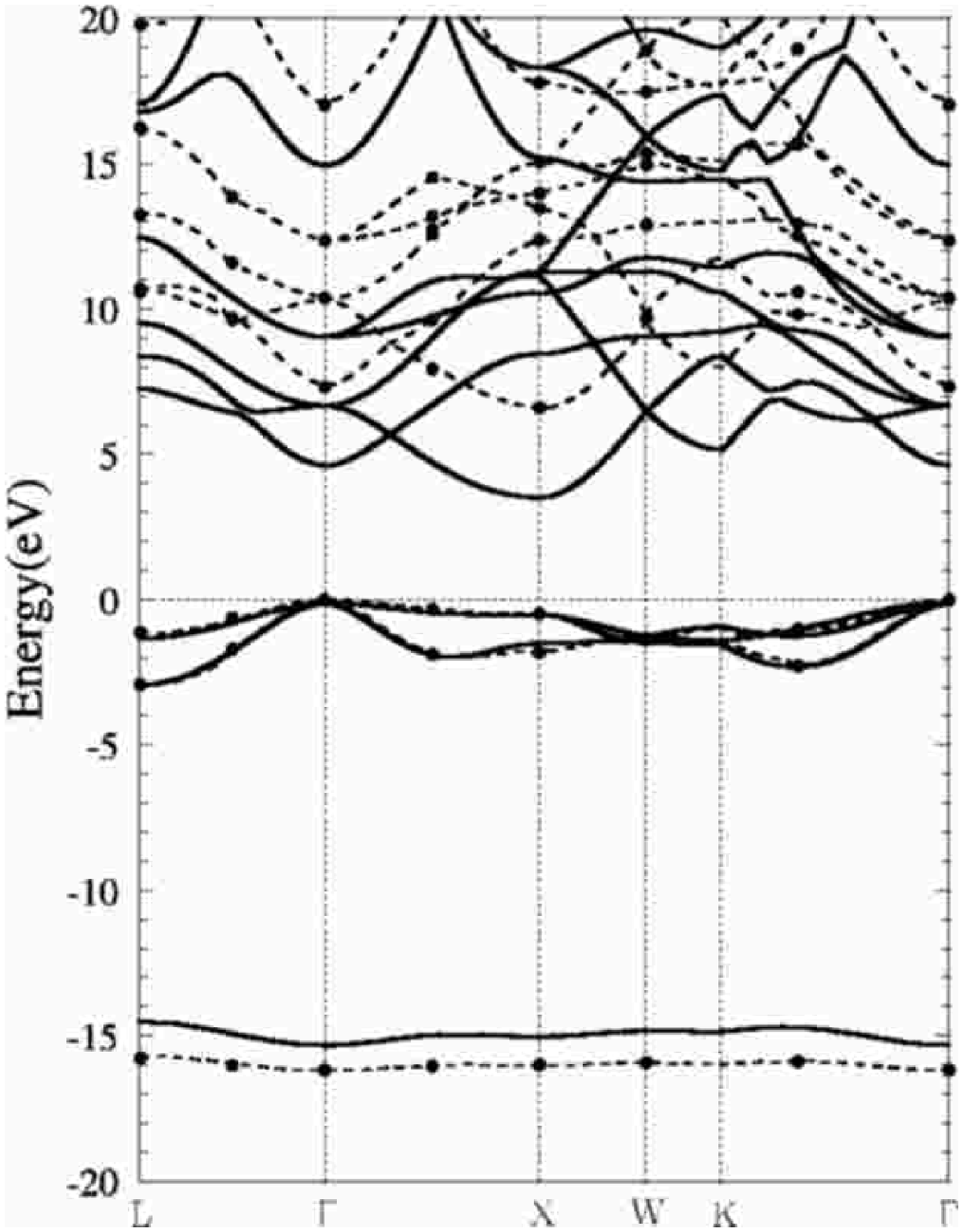}
\end{center}
\caption{The energy bands of MgO (left) and CaO (right), 
         calculated by the LDA (solid lines) 
         and the GWA  (dashed lines) along high symmetry lines. 
         The closed circles are the calculated points in the GWA.
         The high symmetry points are 
         $L$=$(1/2,1/2,1/2)$,
         $\Gamma$=$(0,0,0)$, $X$=$(1,0,0)$, $W$=$(1,1/2,0)$, and 
         $K$=$(3/4,3/4,0)$.}
\label{MgOband}
\end{figure*}%

The lattice constants of MgO, CaO, TiO, and VO are 
$a=$ 4.2122, 4.8105,  4.1766,
and 4.062~\AA, respectively.~\cite{wyckoff}
Empty spheres (ES's) are often necessarily introduced in the ASA
formalism except in closed packing lattices.
The calculation of insulating MgO and CaO without ES gives a
larger GW band gap by about 1~eV than that of calculation with one
ES. 
In the metallic TiO and VO, proper states are not obtained by the 
calculation without ES's in the LDA.
In the case of two or more ES's in NaCl type structure, the radii of ES
become too small.
In the calculation with two ES's the band gap is almost the same as in
the calculation with one ES.
Therefore, the one type of ES is enough in the NaCl type structure.
In the present paper, we use the following sets of radii (in atomic
unit) of atomic spheres; 
(Mg, O, ES)=(2.21, 2.21, 1.62), 
(Ca, O, ES)=(3.03, 2.19, 1.49),
(Ti, O, ES)=(2.71, 1.87, 1.13), and
(V,  O, ES)=(2.58, 1.91, 1.12).
In metallic systems we use 512 ($8\times8\times8$) mesh points to have
enough  convergence.
On the other hand,
in insulating systems smaller number of mesh points 
64 ($4\times4\times4$) is good enough 
to achieve a convergence because of the existence of the band gap.

We must treat the wide energy range 
in the calculation of the dielectric function  and, therefore,  
use a sufficient number of LMTO's. 
The energy-dependent polarization function $\chi^0(\omega)$ has
a long tail in higher energy range and, therefore,
metal $f$ and oxygen $d$ or $f$ orbitals are included for the GWA
in the present work, though these orbitals are omitted usually in the
LDA.
The existence of ES is important
but higher order orbitals in ES are not essential.
The set of the maximum orbitals $l$ of the LMTO basis in 
M, O and ES atomic spheres are chosen to be 
$(ffs)$ for MgO,
$(ffd)$ for CaO,
$(fds)$ for TiO, and 
$(fds)$ for VO.
In CaO, since the band gap becomes narrower, we include the basis of ES
with higher angular momentum.
There are no essential differences between the different choices 
but the above choice is the best for the band gap within 
$\pm0.1~{\rm eV}$.

\section{Results and Discussions} \label{sec3:result}
\subsection{Energy band structure}
\subsubsection{Insulator : {\rm MgO} and {\rm CaO}}
MgO and CaO are ionic crystals and 
are band insulators. 
In MgO, the band gap separates the unoccupied oxygen 2$p$ valence bands 
and the unoccupied magnesium 3$s$-3$p$ conduction bands.
In CaO the band gap locates between the unoccupied 4$s$-4$p$ conduction 
bands and the unoccupied calcium 3$d$ conduction bands.
The band structures of MgO and CaO, calculated both in the LDA
and the GWA, are shown in Fig.~\ref{MgOband} along 
high symmetry lines. 
The band gap, the width of the upper valence bands and the position of 
the low lying oxygen $2s$ band are summarized in Table~\ref{MgOCaOtbl}, 
in comparison with those by the LDA and the HF approximation.
Our results of MgO are in good agreement with those of the previous 
GW calculation.~\cite{GW-MgO}

The band gap in the GWA is found to be 8.2~eV in MgO
and 6.64~eV in CaO, which are in good agreement with the experimental
values 7.8 and 7.1~eV,~\cite{MgO-CaO_exp} respectively.  
The band gap in MgO is the direct one at $\Gamma$ point 
both in the LDA and the GWA as in experiments. 
The band gap in CaO is the indirect one between  $\Gamma$ and $X$
points in the GWA as in the LDA, although the energy difference 
between the $\Gamma$-$\Gamma$ gap and
$\Gamma$-$X$ gap becomes smaller in the GWA. 
The conduction band minimum at the $X$ point originates from metallic $d$ 
states and the lowering the conduction band minimum at the $X$ point
is due to deepening of the potential of the metal ion. 
On the other hand, the direct gap of CaO at the  $\Gamma$ point 
was proposed by the HF approximation 
with correlation calculated by the second order 
perturbation theory.~\cite{HFcorr} 
We are not aware of any angle-resolved
photoemission or inverse photoemission experiments on CaO, and the
nature of the fundamental gap of CaO cannot be concluded at present.

The GWA valence band width is 5.0~eV in MgO and 2.9~eV in CaO,
which are in good agreement with experiments 
and almost the same as those of the LDA. 
The LDA valence band width is generally narrower in general than 
the GWA and the observation, 
because the wave functions of valence states are more
concentrated on atomic sites by local approximation in the LDA,
and the effect of the exchange and correlation
effects are much enhanced.~\cite{Bwidth} 
However, because the valence bands are oxygen 2$p$ states, 
such exchange and correlation effects do not affect much 
in MgO and CaO and band width reduction is not much.
In the calculation with the exact-exchange (EXX) 
potential,~\cite{EXX}
the valence band width of MgO and CaO is narrower than the one in the
LDA but this is due to the different reason. 
In the EXX calculation, the exchange energy  is approximated  by the 
spherically symmetrized Fock term and
the correlation energy is replaced by that of the LDA.
The EXX potential is larger in comparison with the LDA exchange potential 
and gives narrower valence bands of MgO and CaO.
In the HF approximation,  which includes the non-local interaction, 
the valence band width is broadened, for instance, in
MgO.~\cite{MgO-HF1,MgO-HF2} 
But the band gap and band width are overestimated in
the HF approximation due to the lack of screening effects.
In the GWA the valence band width is reduced in comparison with
the HF method.

\begin{table*}
\caption{The band width of O 2$p$ band $W_{2p}$, 
         the position of O $s$ band $E_{O_s}$,
         and the band width of transition metal 3$d$ band $W_{3d}$, 
         $W_{t_{2g}}$, and $W_{e_g}$ 
         in metallic TiO and VO.
         The unit is (eV).}
\begin{ruledtabular}
\begin{tabular}[t]{lcccccccc}
             & \multicolumn{3}{c}{TiO}&
             & \multicolumn{3}{c}{VO}\\ 
             & LDA      & COHSEX      & GWA       &
             & LDA      & COHSEX      & GWA         \\ \hline 
$W_{2p}$     & 4.6         & 4.7         & 4.1         & 
             & 5.2         & 5.3         & 4.6         \\ 
$E_{O_s}$    & $\sim$ 22   
             & $\sim$ 25   & $\sim$ 23   & 
             & $\sim$ 22.5 & $\sim$ 25   & $\sim$ 23   \\ \hline
$W_{3d}$     & 8.4         
             & 9.1         & 6.8         &
             & 8.1         & 7.6         & 6.4         \\
$W_{t_{2g}}$ & 6.5         
             & 6.1         & 4.3         & 
             & 6.4         & 5.4         & 3.8         \\
$W_{e_g}$    & 4.1        
             & 4.1         & 3.2         & 
             & 4.5         & 4.4         & 3.3         \\ 
\end{tabular}
\end{ruledtabular}
\label{TiO3dtbl}
\end{table*}%

The position of oxygen 2$s$ band is pushed down in the GWA
so it is close to the experimental value.
The LDA makes  the core or semicore states locate 
into shallow energy regions, compared with experiments, 
because the LDA potential contains self-interaction. 

\begin{figure}[b]
\begin{center}
 \includegraphics[width=42mm,clip]{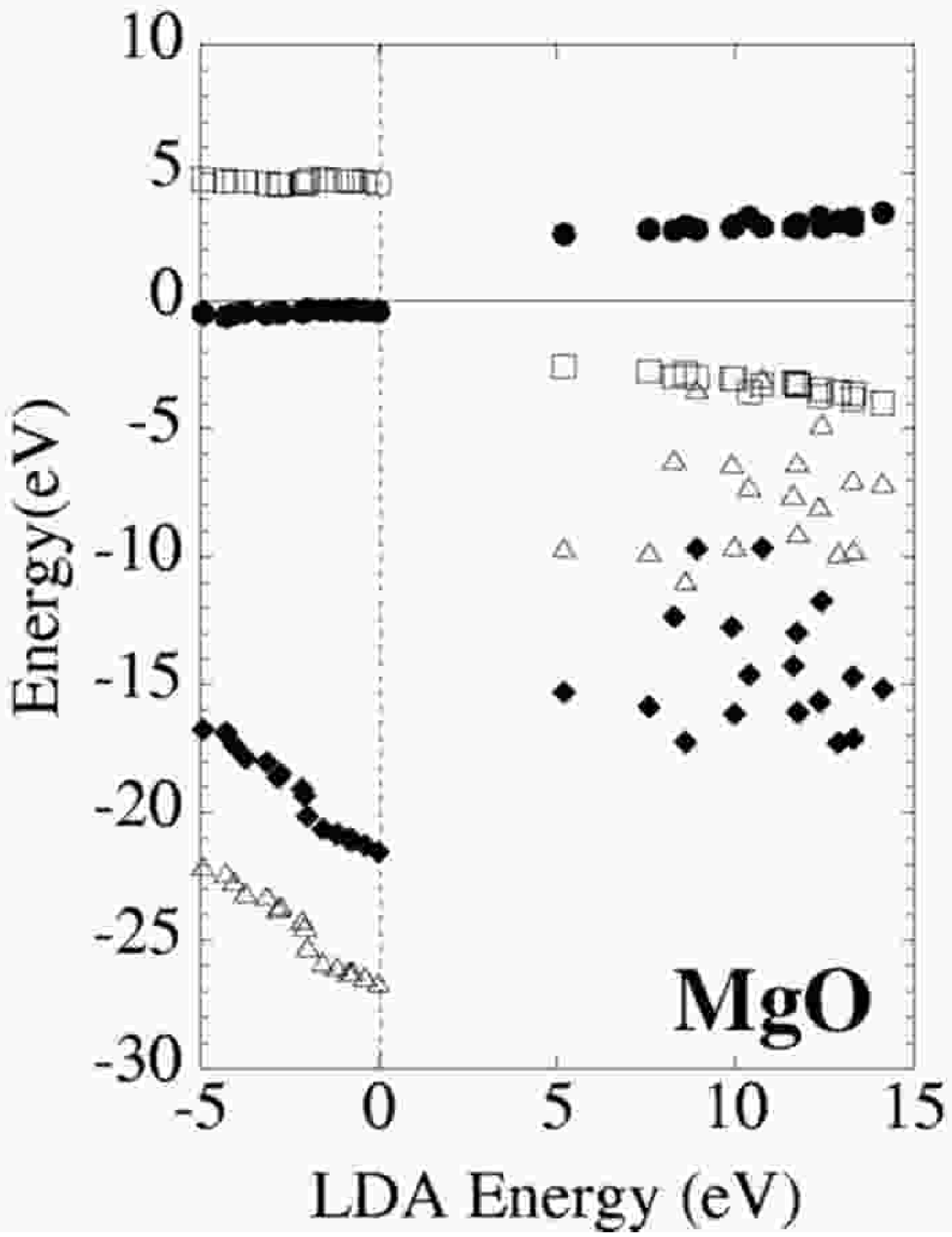}
 \includegraphics[width=42mm,clip]{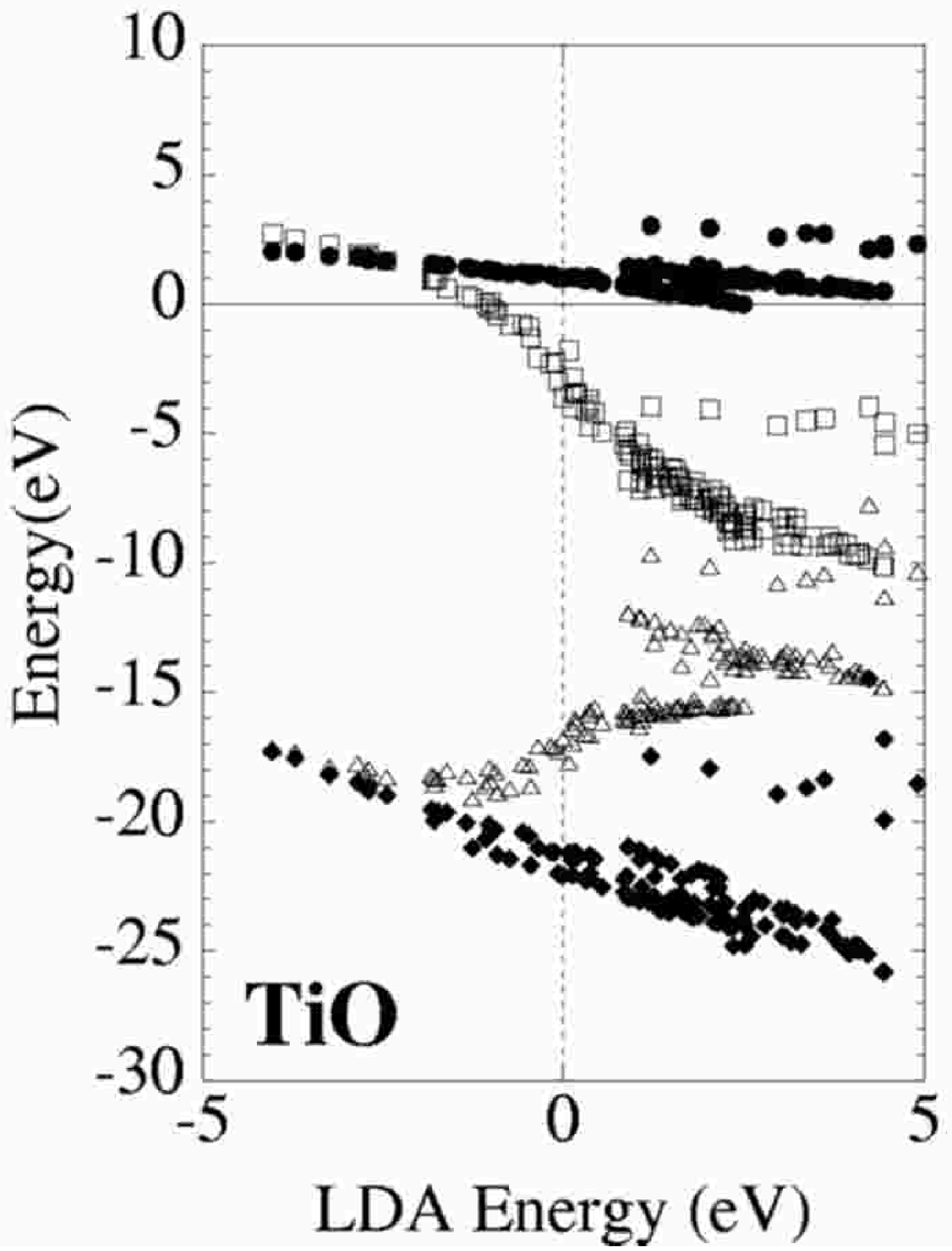}
\end{center}
\caption{Energy dependence of the self-energy of MgO (left) and 
         TiO (right).
         The closed diamond for $V^{\rm xc}_{\rm LDA}$,
         the open triangles for $\Sigma^x$,
         the open squares for $\Sigma^c$, and
         the closed circles for $Z\Delta \Sigma$.}
\label{InsSelE}
\end{figure}

The self-energy and the correction $Z\Delta\Sigma$ 
to the LDA energy of MgO are shown in 
the left side of Fig.~\ref{InsSelE}. 
The gap at the top of the valence bands (the energy zeroth) is seen 
in the self-energy 
and as a result the energy correction $Z\Delta\Sigma$ has the discontinuity.
In conduction bands, fluctuations of $\Sigma^x$ and
$V^{\rm xc}_{\rm LDA}$ are caused by different orbital characters, but
the correction terms $Z\Delta \Sigma$ converge to almost constant value.
The correction terms of the highest valence band $Z_{\rm HVB}$ and 
those of the lowest conduction band $Z_{\rm LCB}$ are as follows:
in MgO, $Z_{\rm HVB}$ = $0.75$--$0.8$ and 
$Z_{\rm LCB}$ = $0.83$--$0.87$; 
in CaO, $Z_{\rm HVB}$ = $0.73$--$0.75$ and 
$Z_{\rm LCB}$ = $0.80$--$0.85$. 
The smaller $Z$ of CaO shows that the electron-electron correlation in
CaO is stronger than in MgO.

To see the dynamical screening effects in MgO and CaO, 
we also calculated the energy band structure by 
using the static COHSEX approximation and shown in 
Table~\ref{MgOCaOtbl}. 
In the static COHSEX approximation, the band gap is much smaller 
than the HF results but is still overestimated to be 9.6~eV
in MgO and 7.7~eV in CaO. 
The reduction of the band gap from the HF result to the static COHSEX
result and that from the static COHSEX result to the GW result are
attributed to the static screening effect and the dynamical screening
effect, respectively.
Therefore, in MgO and CaO, the static screening is the major contribution 
rather than the dynamical effects. 
The valence band width is 5.6~eV in MgO, and 3.4~eV in CaO 
in the static COHSEX approximation.
In the GWA the valence band width is slightly reduced in comparison with
the static COHSEX approximation. 
The reduction from that of the HF approximation to that 
of the static COHSEX approximation is much larger in MgO, because 
of the large  difference of electronegativity and 
resultant importance of the static screening. 
The comparison of the band width by the GWA with that by 
the static COHSEX approximation clearly shows 
that the dynamical effects are appreciable 
in the electron-electron correlation. 

 
\begin{figure*}
\begin{center}
 \includegraphics[width=75mm,clip]{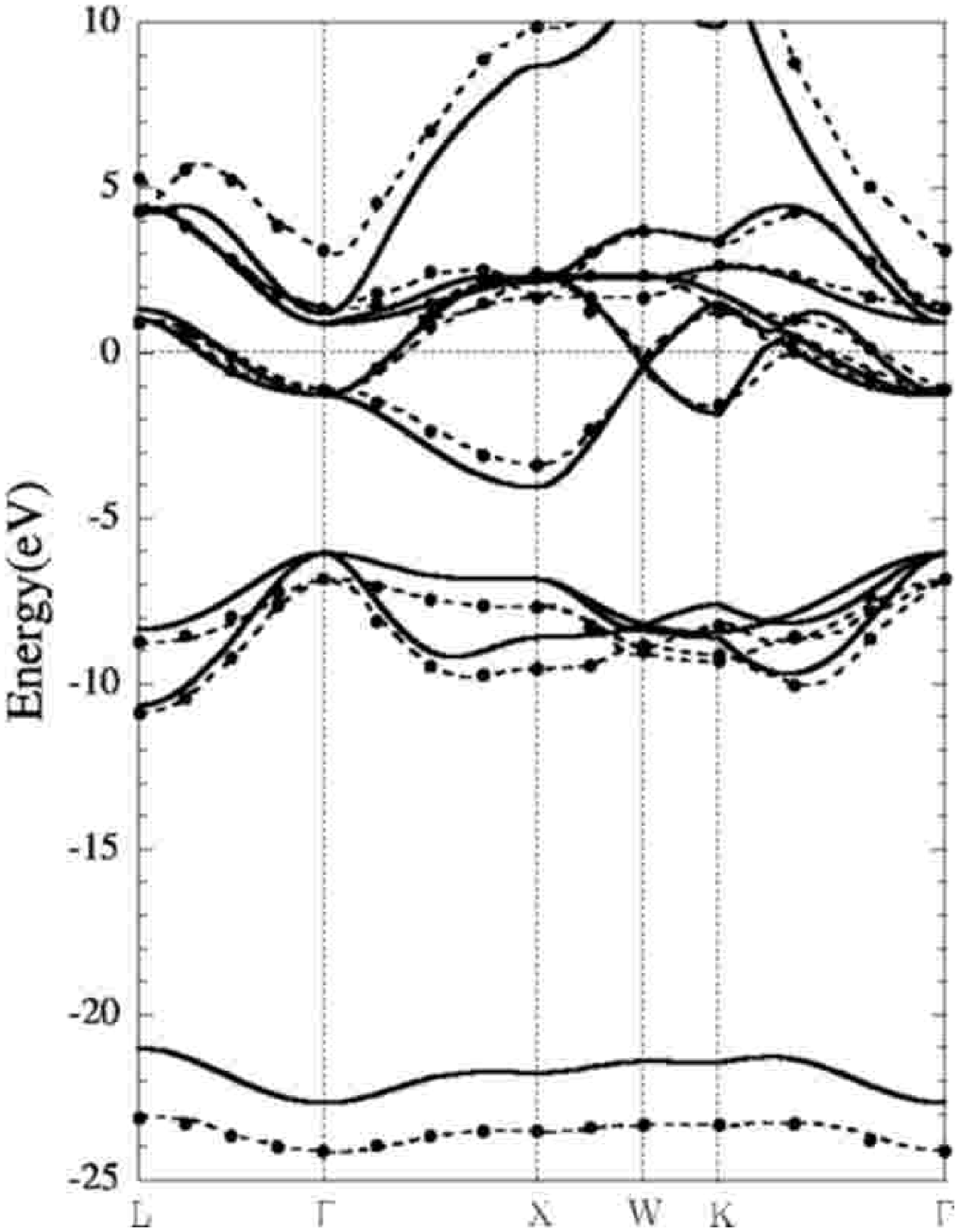}
 \hspace{5mm}
 \includegraphics[width=75mm,clip]{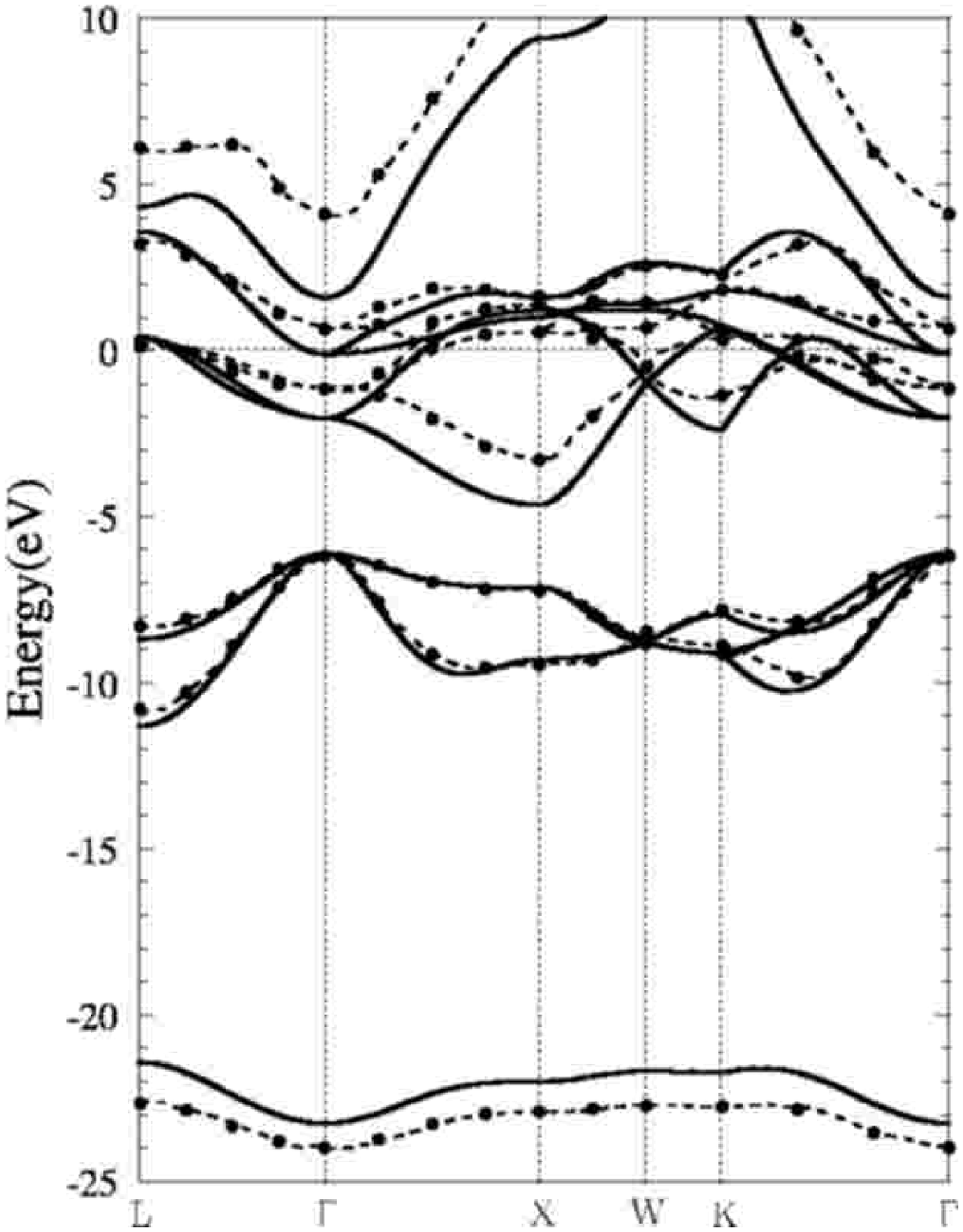}
\end{center}
\caption{The energy bands of TiO (left) and VO (right), 
         calculated by the LDA (solid lines) and by the GWA (dashed lines) 
         along high symmetry lines.  
         The closed circles are the calculated points in the GWA.
         The highest energy band  in the figure is mainly from metal $s$
         states. 
         For the symmetry points, see the caption of  Fig.~\ref{MgOband}.}
\label{TiOband}
\end{figure*}%

\subsubsection{Metal : {\rm TiO} and {\rm VO}}
TiO and VO are metals, showing 
typical properties of covalent crystals such as high melting 
temperature and extreme hardness.
The band structures of TiO and VO are shown in Fig.~\ref{TiOband},
calculated both in the LDA and the GWA and summarized 
in Table~\ref{TiO3dtbl}. 
In the LDA the oxygen 2$s$ states are located around about 22~eV below
$E_F$ in TiO and VO.
These oxygen $2s$ states are separated by about 11~eV from 
the bonding states consisting of metal 3$d$ and oxygen 2$p$ orbitals 
in both TiO and VO, mainly oxygen 2$p$, 
at $7\sim 10~{\rm eV}$ below $E_F$.
A real gap of about 2~eV in TiO and that of about 1.5~eV in VO  can be seen  
between this  bonding states (mainly oxygen 2$p$) 
and the antibonding states (mainly transition metal 3$d$) at
around $E_F$.
The occupied conduction states are heavily dominated by
transition metal 3$d$ orbitals.

The band width of metal 3$d$ states becomes narrower in the GWA
than in the LDA.  
The occupied 3$d$ band width from the UPS
and XPS is about 3~eV in TiO$_{1.03}$.~\cite{TiOexp}
The width of the unoccupied 3$d$ band in the GWA agrees with 
that of the BIS spectra.~\cite{TiOexp}

In TiO and VO the  width of oxygen 2$p$ valence band  in the GWA becomes
narrower in comparison with that of the LDA, which 
is not the case in MgO and CaO. 
The hybridization is very small in MgO and CaO. 
In TiO and VO, the hybridization between the transition
metal 3$d$ and oxygen 2$p$ increases 
due to deepening metal 3$d$ states with the increase of the $d$
electron.  
Once one adopts the GWA, the wave functions of valence bands are
localized more due to the screening effects 
and the width of the band around $-10$~eV becomes narrower. 
Then the conduction bands are lifted to higher 
energy side, relative to the lower energy oxygen $2p$ (bonding) band
in TiO and VO. 
The effects of the reduction of the conduction (transition metal $3d$)
band width also causes the increasing the separation between  $e_g$ and
$t_{2g}$ states. 
The position of oxygen 2$s$ band is also  pushed down by the GW correction,
and this is caused by the self-interaction correction as in MgO and
CaO. 

The self-energy and the correction $Z\Delta\Sigma$ to the LDA energy 
of TiO are shown in the right side of Fig.~\ref{InsSelE}. 
In contrast to insulating system MgO
three parts of the self-energy and the
energy correction $Z\Delta\Sigma$ are continuous at the Fermi energy
$E_F$, which are the
typical properties of the Fermi liquid. 
The renormalization factor of transition metal 3$d$ states is 
$Z$ = $0.55$--$0.65$ in TiO and $Z$ = $0.51$--$0.63$ in VO.
$Z$ of oxygen 2$p$ states is about $0.7$--$0.8$ in both systems. 
Those results of the renormalization factor suggest that 
the interaction between 3$d$ electrons is strong,
and the correlation in VO is larger than in TiO.

\begin{figure}[b]
\begin{center}
 \includegraphics[height=57mm,clip]{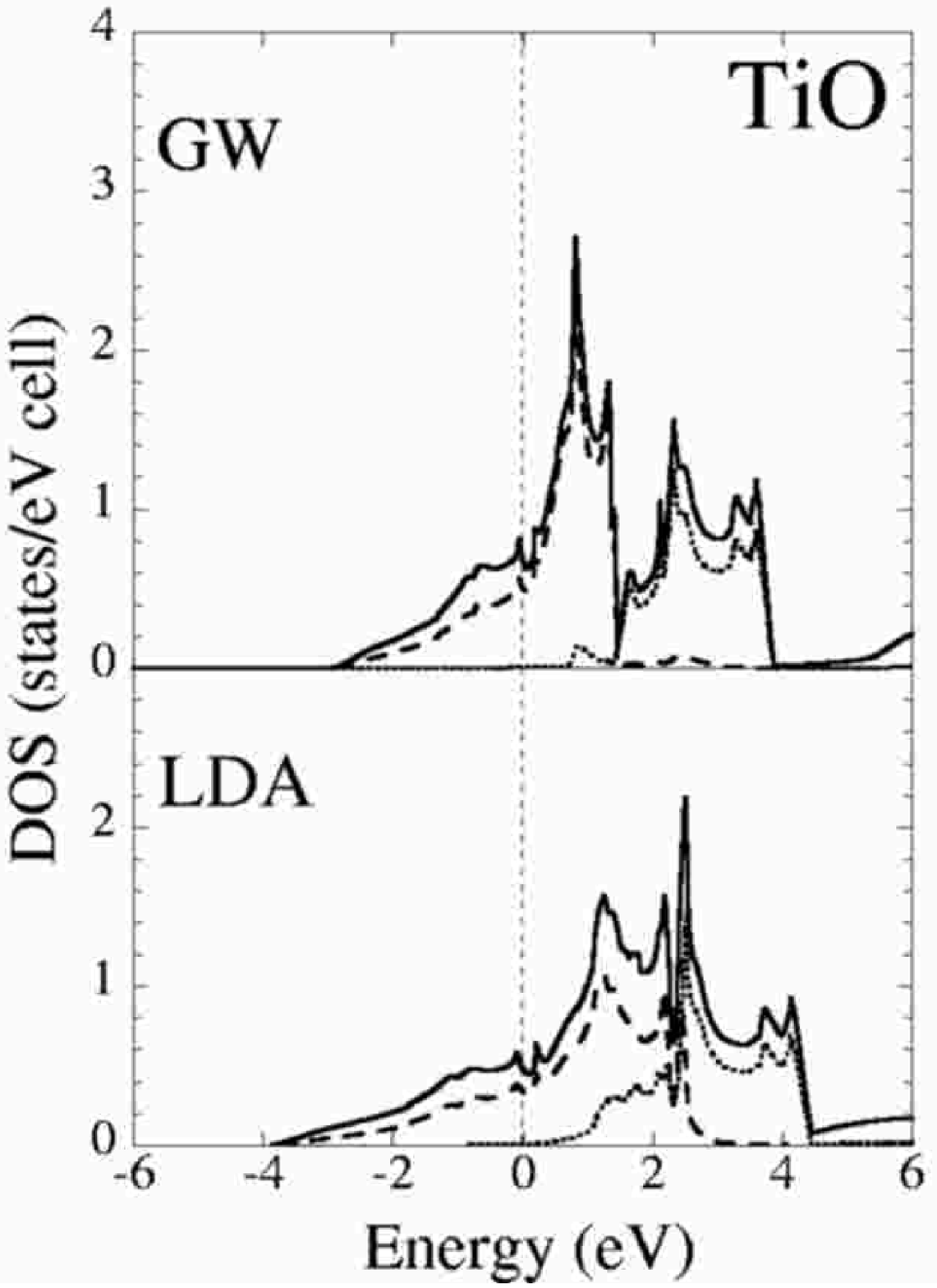}
 \includegraphics[height=57mm,clip]{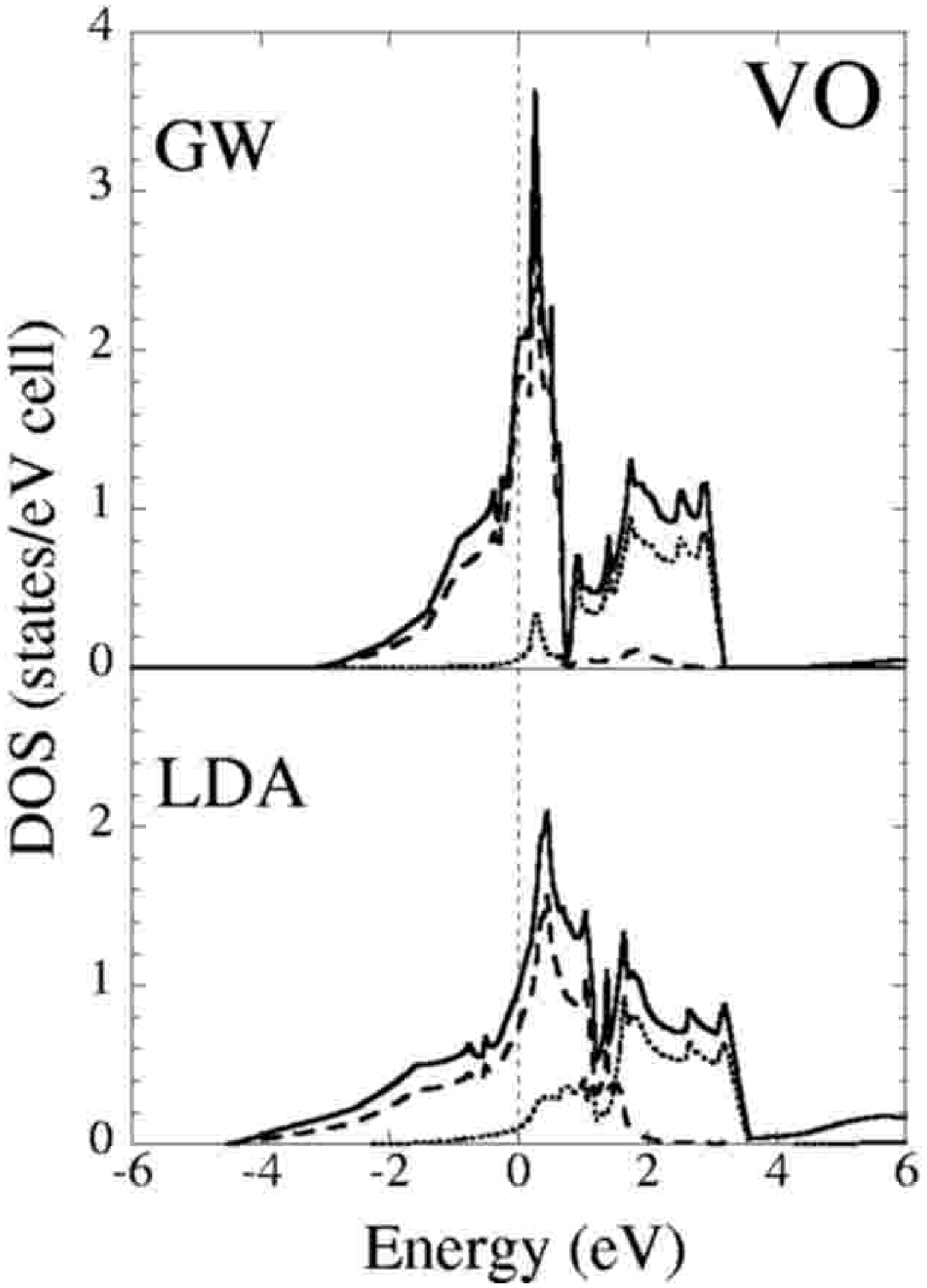}
\end{center}
\caption{Total and projected DOS in a transition metal atom 
         in TiO and VO by the LDA (bottom) and the GWA (top).
         The solid lines for the total DOS in a unit cell, 
         the dotted lines for the partial DOS of transition metal
         3$d$-$e_g$ states and 
         the chain lines for the partial DOS of  transition metal
         3$d$-$t_{2g}$ states.  
         The contribution by the oxygen atom is small in this energy
         range.}
\label{TiOdos}
\end{figure}%

\begin{figure*}
\begin{center}
 \includegraphics[width=179mm,clip]{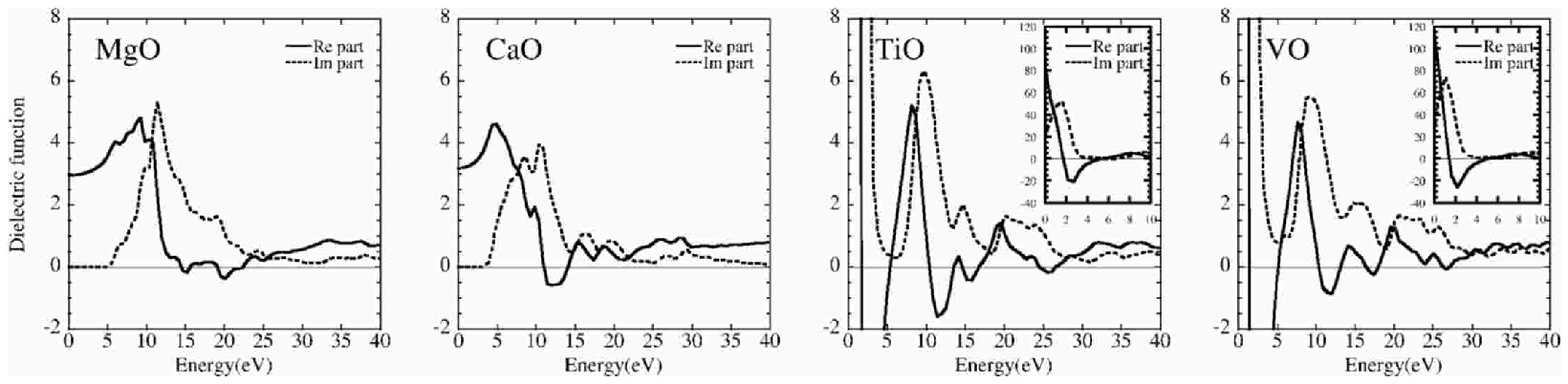}\\
 \includegraphics[width=179mm,clip]{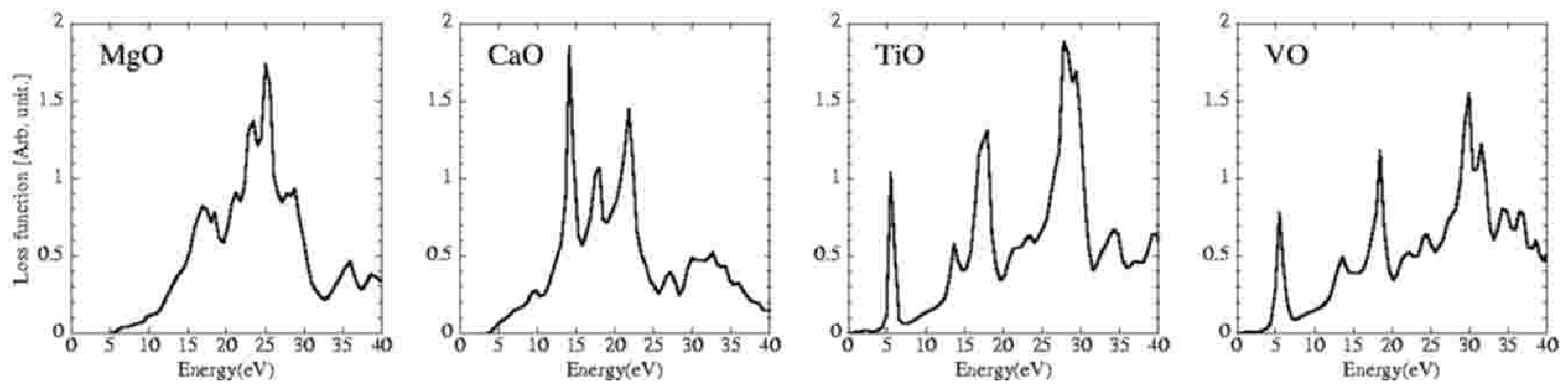}
\end{center}
\caption{From left to right, spectra of MgO, CaO, TiO, and VO are shown. 
         Upper panel and lower panel are the dielectric function
         $\varepsilon(\omega)$ and the electron energy-loss spectra
         (EELS) $\varepsilon^{-1}(\omega)$ for ${\bf q}= 2\pi/a(0.25,0,0)$.}
\label{EELS}
\end{figure*}%

The $d$-$d$ Coulomb interaction
$\langle \phi_d \phi_d |W(\omega =0)|\phi_d \phi_d \rangle$ 
is calculated to be about 1.5~eV both in TiO and VO. 
The bare Coulomb interaction
$\langle \phi_d \phi_d |v|\phi_d \phi_d \rangle$ 
is  actually $15.3$~eV in TiO and 
$17.8$~eV in VO. 
Then the correlation term
$\langle \phi_d \phi_d |W^c(\omega =0)|\phi_d \phi_d \rangle$ 
is  $-13.7$~eV in TiO and $-16.3$~eV in VO. 
Therefore, the correlation effects and the static screening 
(at the Fermi energy) are quite important in these metallic oxides.
We should mention that the $d$-$d$ Coulomb interaction 
$\langle \phi_d \phi_d |W(\omega =0)|\phi_d \phi_d \rangle$ 
is different from the Hubbard $U$ evaluated from 
the LDA, which includes only screening by on-site $d$ electrons. 
The term  
$\langle \phi_d \phi_d |W(\omega =0)|\phi_d \phi_d \rangle$  
includes the screening effects by both on-site and 
off-site $d$ electrons.~\cite{Aryasetiawan-W} 

The density of states (DOS) of transition metal 3$d$ orbitals 
are shown in Fig.~\ref{TiOdos}. 
In the calculation of the density of states in  
the present GWA,  
the eigenvalues were replaced by those of the GWA 
but the eigenfunctions were the same as those of the LDA, 
and the lifetime (from imaginary part of the self-energy) of
quasiparticle was not considered.
The transition metal $e_g$ states are pushed up in the GWA 
and the $e_g$-$t_{2g}$ separation becomes much clearer than in LDA results. 

To see the dynamical correlation effects in TiO and VO 
we also show the results by the static COHSEX approximation 
in Table~\ref{TiO3dtbl}. 
The band width of the static COHSEX approximation does not become
narrower appreciably  in comparison with the LDA results.
The reduction by the GWA is much appreciable.  
We could conclude that the band width reduction is mainly 
due to the dynamical correlations.
The importance of the dynamical correlation 
was also mentioned in the 3$d$ band width of 
$3d$ ferromagnetic transition metals.~\cite{QP}

Application of the LDA+U method to 
TiO and VO lead to antiferromagnetic insulators rather 
than paramagnetic metals, when one uses the $U$ value 
obtained by the LDA and the photoemission spectra.~\cite{LDA+U1}
However, as already shown, the $d$-$d$ Coulomb interaction 
$\langle \phi_d \phi_d |W^c(\omega =0)|\phi_d \phi_d \rangle$ 
is much smaller by the correlation effects. 

\subsection{Energy loss spectrum and plasmon frequency}
The dielectric function is seen in the procedure 
of the GW calculation. 
We show the dielectric functions and the electron energy-loss spectra
(EELS) in Fig.~\ref{EELS} for ${\bf q}= 2\pi/a(0.25,0,0)$,
using  512 ${\bf k}$ points in the Brillouin zone.  

The dielectric function consists of a part of free
electron contribution and a part of interband
transition.~\cite{dielectric}
The plasmon frequency $\omega_{\rm pl}$ corresponds 
to the zeroth of the free electron part of the 
dielectric function and the peak of the EELS. 
When the interband transition exists at lower energy side 
of the plasmon frequency, 
the frequency of vanishing $\varepsilon({\bf q},\omega)$ shifts 
towards higher frequency side.
This is the shift of the plasma edge in the dielectric 
functions of metals or the shift of the plasmon peak in the EELS. 
The peaks of the imaginary part of the dielectric function 
$\varepsilon(\omega )$ appear essentially 
at the energies of interband excitations.
On the other hand, the peaks of the EELS appear at positions 
where the dielectric function vanishes or becomes 
very small. 
Therefore, the EELS have complex structure originated 
from the situation of the coexistence, in a narrow energy region, 
of the plasma edge and the interband transition. 
This is the case in transition metal monoxides  and  
it may be very interesting to see the difference or  
the similarity of the dielectric functions and the EELS 
of MgO, CaO, TiO, and VO.

The interband transition from O 2$p$ to metal 3$d$ appears at
about 10~eV in CaO, TiO, and VO  but that of  MgO locates  
at much higher energy.
In MgO, the transition from O 2$p$ to metal 3$s$-3$p$ 
creates a broad peak of $\varepsilon(\omega )$  at about 12~eV. 
On the other hand, the transition from O 2$p$ to metal 4$s$-4$p$
makes a  structure of $\varepsilon(\omega )$
at about $15\approx20$~eV in CaO, at about
$20\approx27$~eV in TiO and VO.
The sharp peaks of  $\varepsilon(\omega )$
at 1.5~eV in TiO and at 1.1~eV in VO are created  by metal
3$d$-3$d$ transition.
The dielectric function of metals should 
reduce generally to zero near $\omega =0$ 
and then show the Drude peak at $\omega =0$.
In metallic compounds TiO and VO, low energy excitation peaks 
of  $\varepsilon (\omega )$ exist down to $0.2\approx0.3$~eV.
Then the nonzero value of $\text{Im} \varepsilon(\omega=0)$ 
in Fig.~\ref{EELS}  is the artifact caused 
by broadening these low frequency peaks 
of  the irreducible polarization function  $\chi^0$ 
by Gaussian with a width of $\sigma \approx 0.7$~eV. 

In the EELS, the sharp peak of metallic TiO and VO at 5~eV is originated
from metal 3$d$-3$d$ transition.
The transition from O 2$p$ to metal 3$d$ generates the peak at 14~eV in
CaO and at 18~eV in TiO and VO.
The shoulder at around 17~eV in MgO is caused by the transition from O 2$p$
to metal 3$s$-3$p$ because the imaginary part of dielectric function has
a relatively large nonzero value at this energy due to broad peak of
interband transition.
The plasmon peak shifts 
from  21.1 to 25.0~eV in MgO, 
from  17.2 to 21.8~eV in CaO, 
from  24.6 to 27.8~eV in TiO, and 
from  27.2 to 29.9~eV in VO 
due to interband transitions. 
In this estimation of the plasmon frequency,  
we adopted the number of electrons to be 
6 for MgO and CaO (metal $s$ and O 2$p$ electrons),  
8 for TiO (Ti 3$d$,4$s$ and O $2p$), and 
9 for VO (V 3$d$,4$s$ and O $2p$). 

\section{Summary} \label{sec4:summary}
We applied the GWA to the nonmagnetic oxides MgO, CaO, TiO, and VO.
Band gap and occupied oxygen 2$p$ band width in insulating MgO and
CaO are in good  agreement with  experimental results. 
In both systems oxygen 2$s$ level is pushed down, 
because the self-interaction is corrected in the GWA.
We compare the results by the GWA with those by the 
static COHSEX approximation and discussed 
the effects of the dynamical correlation. 
In metallic TiO and VO, the band width of transition metal $e_g$ and
$t_{2g}$ states become narrower due to the correlation effects. 

We also present values of the $d$-$d$ Coulomb interaction 
in TiO and VO  
and showed the crucial role of the dynamical correlation in these
systems. 
The GW method may be one of the most appropriate method 
which can provide the information of the electron-electron interaction 
with screening effects. 

The actual wave functions should differ from those in the
LDA when the effects of dynamical correlation are included. 
The choice of the starting wave functions or equivalently the choice of
the unperturbed Hamiltonian in the GWA may be 
very important from this point.  
This may be very essential for the characteristics 
in the magnetic transition metal oxides 
but still open for the future study.  

\begin{acknowledgments}
We thank to F. Aryasetiawan and T. Kotani for
useful discussions and suggestion about our GW code development.
This work is supported by Grant-in-Aid for COE Research
``Spin-Charge-Photon'', and Grant-in-Aid from the Japan Ministry of
Education, Science, Sports and Culture. 
Part of the present calculation has been done by use of the facilities
of the Supercomputer Center, Institute for Solid State Physics,
University of Tokyo.
\end{acknowledgments}

\appendix
\section{Computational details}\label{App.A}
Since the wave function $\psi$ is written in terms of 
the LMTO basis functions $\chi$,
the Hilbert space of the self-energy is spanned by the product basis as
\begin{equation}
 \{ \Sigma \}=\{ \psi \psi \}=\{ \chi \chi \}
  = \{ \phi \phi \}+\{ \phi {\dot \phi} \}+
  \{ {\dot \phi} {\dot \phi} \}    \ .
\end{equation}
In the present calculations, 
because of a proper choice of the energy parameters
the absolute values of the mixing 
coefficients of $\dot{\phi}$ to $\phi$ are less than $0.1$  for all terms 
and the norm of $\dot{\phi}$, $\langle \dot{\phi}\dot{\phi} \rangle$,  
is $0.1\sim 0.3$ even in the largest case.
Then the terms including $\dot \phi$ can be dropped out safely  
and the self-energy $\Sigma$ can be expanded by 
$\phi_{l^\prime m^\prime \nu^\prime}^*({\bf r})\phi_{lm\nu}({\bf r})$ = 
$\phi_{l^\prime\nu^\prime}(|{\bf r}|)\phi_{l\nu}(|{\bf r}|)
Y_{l^\prime m^\prime}^*(\hat{\bf r})Y_{lm}(\hat{\bf r})$.
The product of two spherical harmonics is reduced to the linear
combination of one spherical harmonics.
Consequently we can construct the new basis set
\begin{align}
 & \tilde{B}_{{\bf R}l^\prime \nu^\prime l\nu;
 l^{\prime\prime}m^{\prime\prime}}
 ^{\bf k} ({\bf r}_{\bf R})
 \notag\\
 & = \sum_{\bf T} e^{i{\bf k}\cdot{\bf T}}
 \phi_{l^\prime \nu^\prime}(|{\bf r}_{{\bf R}{\bf T}}|)
 \phi_{l\nu}               (|{\bf r}_{{\bf R}{\bf T}}|)
 Y_{l^{\prime\prime}m^{\prime\prime}}
 (\widehat{{\bf r}_{{\bf R}{\bf T}}}) ,
\end{align}
where $L^{\prime\prime}$ is depend on $l^{\prime}$ and $l$, that is
$|l-l^{\prime}| \leq l^{\prime\prime}(l^\prime,l) \leq l+l^{\prime}$ and 
$|m^{\prime\prime}(l^\prime,l)| \leq l^{\prime\prime}(l^\prime,l)$, and 
${\bf r}_{{\bf R}{\bf T}} = {\bf r}-{\bf R}-{\bf T}$.
The orthonormal basis set $B$ can be produced by the linear combination
of the nonorthogonal basis set $\tilde B$ as 
$B_i^{\bf k}=\sum_j \tilde{B}_j^{\bf k} D_{ji}^{\bf k}$, where
$j=({\bf R}l^\prime \nu^\prime l\nu;l^{\prime\prime}m^{\prime\prime})$,
$i$ is a new suffix in the orthonormal basis set and $D$ is the 
transformation matrix from the nonorthonormal basis to the
orthonormal basis. 
Here we use the convenient notation
\begin{equation}
 |B_i^{\bf k} \rangle \equiv |i_{\bf k} \rangle 
  = | \tilde{i}_{\bf k} \rangle  D^{\bf k} ,  \ \ 
 | \tilde{i}_{\bf k} \rangle \equiv |{\tilde B}_i^{\bf k} \rangle .
\label{trans-basisA}
\end{equation}
The matrix $D^{\bf k}$ is obtained by orthogonalizing the overlap
matrix $O_{ij}^{\bf k}=\langle \tilde{i}_{\bf k} | \tilde{j}_{\bf k}
\rangle$. 

We can expand the self-energy in terms of the new orthonormal basis
set $| i_{\bf k} \rangle$. 
To calculate the self-energy we first
calculate the polarization function $\chi^0$ with the
nonorthonormal basis.

The Coulomb potential is calculated by using the bare structure constant
of the LMTO method.~\cite{Coulomb}
In this method, two parts of the Coulomb matrix
can be calculated separately, on-site and off-site parts, by
using the LMTO bare structure constant $S^0({\bf q})$. 

Next we need to transform from nonorthonormal basis 
$| \tilde{i}_{\bf k} \rangle$ to orthonormal basis 
$| i_{\bf k} \rangle$, that is 
$\langle \tilde{i}|\chi^0|\tilde{j}\rangle \rightarrow
\langle i |\chi^0| j \rangle$ and
$\langle \tilde{i}|v|\tilde{j}\rangle \rightarrow
\langle i |v| j \rangle$.
Within the orthonormal basis we can calculate the response function
$\chi=(1-\chi^0 v)^{-1}\chi^0$ and the screened Coulomb potential
$W^c=v \chi v$. 
Finally the self-energy can be obtained by the 
inverse transformation from the orthogonal basis to the nonorthogonal 
basis, and that from the product basis to the Bloch basis. 

The number of basis set can be reduced by excluding terms of 
higher $l^{\prime\prime}$. 
We have studied the dependence 
on a choice of the maximum $l^{\prime\prime}$'s  
for set of $l^{\prime\prime}$ in (Mg, O, ES) atomic spheres 
being (3,3,0), (4,4,0) and (6,6,0). 
The configuration (6,6,0) corresponds to the full calculation for 
orbital $f$ for metals, $f$ for oxygen and $s$ for ES. 
The total number of product bases of (3,3,0), (4,4,0), and (6,6,0) are
120, 174, and 222, respectively. 
The proper choice may be a set of the maximum $l^{\prime\prime}$'s  
which produces the total number of the product basis nearly equals to 
a half of that in the maximum choice. 
For example, the product basis of (3,3,0) 
may be enough for the set of the $ffs$ Bloch orbitals. 
In present work the full product basis is selected, but
this way of product basis reduction is very effective and powerful for
large systems.  
%

\section{Numerical technique}\label{App.B}
In the calculation of the self-energy, 
the integration  is  performed  over the whole Brillouin zone.  
The Coulomb matrix $v({\bf q})$ has  a singularity  at ${\bf q}=0$  as 
$F({\bf q})=1/|{\bf q}|^2$.  
The integration of $v({\bf q})$ over the Brillouin zone does not 
diverge but special cares are needed not only for the ${\bf q}=0$ term
but for small finite ${\bf q}$.
In the present calculation, the integration over the Brillouin zone
is replaced by the summation over discrete points. 
We use a set of discrete points distributed densely near the 
${\bf q}=0$ point, not on uniform $\bf k$ mesh, in the Brillouin zone. 
For a choice of the discrete points near ${\bf q}=0$,  
we use the offset $\Gamma$-point method.~\cite{offset}
The integration of $F({\bf q})$ over the Brillouin zone can be performed
analytically and  
the offsetted points
$\bf Q$'s are chosen near ${\bf q}=0$ so as to satisfy a relation 
\begin{equation}
 \int_{\rm BZ} F({\bf q}) d{\bf q} = \sum_{\bf Q} F({\bf Q}) + 
  \sum_{{\bf k}\neq 0} F({\bf k}) ,
\end{equation}
where $\bf k$'s are the discrete mesh points of the Brillouin zone.
The choice of the  offsetted points ${\bf Q}$ are taken into
consideration so as to keep the symmetry of the system. 

We tested the offset method in the exchange
energy of the electron gas system and the accuracy has been confirmed
even in case of small number of ${\bf k}$-mesh points. 
The careful treatment of the Coulomb matrix at or near ${\bf q}=0$ 
is very crucial  near the band gap or the Fermi level.



\begin{thebibliography}{99}
\bibitem{LDA} P. Hohenberg and W. Kohn, 
              Phys. Rev. {\bf 136}, B864 (1964);
              W. Kohn and L. J. Sham, 
              {\it ibid.} {\bf 140}, A1133 (1965);
              R. O. Jones and O. Gunnarsson, 
              Rev. Mod. Phys. {\bf 61}, 689 (1989). 
\bibitem{LDA-TMO} K. Terakura, A. R. Williams, T. Oguchi, and
	          J. K\"{u}bler, Phys. Rev. Lett. {\bf 52}, 1830 (1984); 
	          K. Terakura, T. Oguchi, A. R. Williams, and
	          J. K\"{u}bler, Phys. Rev. B {\bf 30}, 4734 (1984).
\bibitem{GGA} Y. Ishii, in 
              {\it Computational Physics as a New Frontier in Condensed
	           Matter Research}, 
              edited by H. Takayama, M. Tsukada, H. Shiba, F. Yonezawa,
	                M. Imada, and Y. Okabe 
              (The Physical Society of Japan, Tokyo, 1995), pp. 57--66.
\bibitem{SIC2} M. Arai and T. Fujiwara,
               Phys. Rev. B  {\bf 51}, 1477 (1995).
\bibitem{SIC3} A. Svane and O. Gunnarsson,
               Phys. Rev. Lett. {\bf 65}, 1148 (1990).
\bibitem{LDA+U1} V. I. Anisimov, J. Zaanen, and O. K. Andersen,
                 Phys. Rev. B {\bf 44}, 943 (1991).
\bibitem{LDA+U2} V. I. Anisimov, I. V. Solovyev, M. A. Korotin, 
                 M. T. Czyzyk, and G. A. Sawatzky,
                 Phys. Rev. B {\bf 48}, 16 929 (1993);
                 A. I. Liechtenstein, V. I. Anisimov, and J. Zaanen,
                 {\it ibid.} {\bf 52}, R5467 (1995).
\bibitem{LDA+U3} V. I. Anisimov, F. Aryasetiawan, and A. I. Lichtenstein, 
                 J. Phys.: Condens. Matter {\bf 9}, 767 (1997).
\bibitem{GW1} L. Hedin and S. Lundqvist, in 
              {\it Solid State Physics}, 
              edited by H. Ehrenreich, F. Seitz, and D. Turnbull 
              (Academic Press, New York, 1969), Vol. 23, P.1.
\bibitem{GW2} F. Aryasetiawan and O. Gunnarsson,
              Rep. Prog. Phys. {\bf 61}, 237 (1998).
\bibitem{QP} M. M. Steiner, R. C. Albers, and L. J. Sham,
             Phys. Rev. B {\bf 45}, 13272 (1992).
\bibitem{Niexp} D. R. Penn, Phys. Rev. Lett. {\bf 42}, 921 (1979);
                A. Liebsch, Phys. Rev. B {\bf 23}, 5203 (1981).
\bibitem{GW-SimpleMetal} J. E. Northrup, 
                         M. S. Hybertsen, and  S. G. Louie,  
                         Phys. Rev. Lett. {\bf 59}, 819 (1987);
                         J. E. Northrup, 
                         M. S. Hybertsen, and S. G. Louie, 
                         Phys. Rev. B {\bf 39}, 8198 (1989).
\bibitem{GW-semicon1} M. S. Hybertsen and S. G. Louie, 
                      Phys. Rev. B {\bf 34}, 5390 (1986).
\bibitem{GW-semicon2} R. W. Godby, M. Schl\"uter, and L. J. Sham, 
                      Phys. Rev. B {\bf 37}, 10 159 (1988).
\bibitem{GW-MgO} U. Sch\"onberger and  F. Aryasetiawan,
                 Phys. Rev. B {\bf 52}, 8788 (1995).
\bibitem{GW-TM} F. Aryasetiawan,
                Phys. Rev. B {\bf 46}, 13 051 (1992).
\bibitem{GW-TMO} F. Aryasetiawan and O. Gunnarsson,
                  Phys. Rev. Lett. {\bf 74}, 3221 (1995).
\bibitem{mGW-TMO1} S. Massidda, A. Continenza, M. Posternak, and
                   A. Baldereschi,
                   Phys. Rev. Lett. {\bf 74}, 2323 (1995).
\bibitem{mGW-TMO2} S. Massidda, A. Continenza, M. Posternak, and
                   A. Baldereschi,
                   Phys. Rev. B {\bf 55}, 13 494 (1997).
\bibitem{mGW-TMO3} A. Continenza, S. Massidda, and M. Posternak,
                   Phys. Rev. B {\bf 60}, 15 699 (1999).
\bibitem{NiO-CI} A. Fujimori and F. Minami,
                 Phys. Rev. B {\bf 30}, 957 (1984).
\bibitem{Conserve} G. Baym and L. P. Kadanoff, 
                   Phys. Rev. {\bf 124}, 287 (1961);
                   G. Baym, {\it ibid.} {\bf 127}, 1391 (1962).
\bibitem{GW-SCF} W.-D. Sch\"{o}ne and A. G. Eguiluz, 
	         Phys. Rev. Lett {\bf 81}, 1662 (1998).
                 In this paper fully self-consistent GW calculation is
	         applied to K and Si, and it gives wider band width in K
	         or overestimated band gap in Si. 
\bibitem{OKA} O. K. Andersen, Phys. Rev. B {\bf 12}, 3060 (1975).
\bibitem{MgO-HF1} S. T. Pantelides, D. J. Mickish, and A. B. Kunz,
                  Phys. Rev. B {\bf 10}, 5203 (1974).
\bibitem{MgO-HF2} T. Bredow and A. R. Gerson,
                  Phys. Rev. B {\bf 61}, 5194 (2000).
\bibitem{HFcorr} R. Pandey, J. E. Jaffe, and A. B. Kunz,
                 Phys. Rev. B {\bf 43}, 9228 (1991).
\bibitem{MgO-CaO_exp} R. C. Whited, C. J. Flaten, and W. C. Walker,
                      Solid State Commun. {\bf 13}, 1903 (1973).
\bibitem{MgO_exp2} S. P. Kowalczyk, F. R. McFeely, L. Ley,
                   V. T. Gritsyna, and D. A. Shirley,
                   Solid State Commun. {\bf 23}, 161 (1977).
\bibitem{pd-basis} F. Aryasetiawan and O. Gunnarsson,
                   Phys. Rev. B {\bf 49}, 16 214 (1994).
\bibitem{wyckoff} R. W. G. Wyckoff, {\it Crystal Structures}
                  (Wiley, New York, 1963).
\bibitem{Bwidth} E. L. Shirley,
                 Phys. Rev. B {\bf 58}, 9579 (1998).
\bibitem{EXX} T. Kotani,
              Phys. Rev. B {\bf 50}, 14 816 (1994); 
	                   {\bf 51}, 13 903(E) (1995);
              T. Kotani and H. Akai,
              {\it ibid.} {\bf 54}, 16 502 (1996).
\bibitem{TiOexp} S. R. Barman and D. D. Sarma,
                 Phys. Rev. B {\bf 49}, 16 141 (1994).
\bibitem{Aryasetiawan-W} M. Springer and F. Aryasetiawan, 
                         Phys. Rev. B {\bf 57}, 4364 (1998).
\bibitem{dielectric} H. Ehrenreich and H. R. Philipp, 
                     Phys. Rev. {\bf 128}, 1622 (1962).
\bibitem{Coulomb} A. Svane and O. K. Andersen,
                  Phys. Rev. B {\bf 34}, 5512 (1986).
\bibitem{offset} T. Kotani and M. van Schilfgaarde
                 (private communications).
\end{thebibliography}
\end{document}